\documentclass[pra,aps,twocolumn,10pt]{revtex4-1}
\usepackage[strict]{revquantum}
\usepackage[bold]{hhtensor}

\definecolor{darkblue}{RGB}{0,0,127} 
\definecolor{darkgreen}{RGB}{0,150,0}
\hypersetup{breaklinks, colorlinks, linkcolor=darkblue, citecolor=darkgreen, filecolor=red, urlcolor=darkblue}
\hypersetup{
	pdftitle={Practical adaptive quantum tomography}, 
	pdfauthor={Christopher Granade, Christopher Ferrie, and Steven T. Flammia}
}

\definelanguagealias{EN}{english}

\newrm{LS}

\newcommand{\figurefolder}{fig}

\begin{document}

\title{Practical adaptive quantum tomography}

\author{Christopher Granade}
\email{cgranade@cgranade.com}
\homepage{http://www.cgranade.com}
\thanks{complete data and source code for this work are available online \cite{supp_mat},
and can be previewed at \href{https://goo.gl/koiWxR}{https://goo.gl/koiWxR}.}

\author{Christopher Ferrie}

\author{Steven T.\ Flammia}

\affiliation{Centre for Engineered Quantum Systems, School of Physics, University of Sydney, Sydney, NSW, Australia}

\date{\today}

\begin{abstract}
We introduce a fast and accurate heuristic for adaptive tomography that addresses many of the limitations of prior methods. 
Previous approaches were either too computationally intensive or tailored to handle special cases such as single qubits or pure states. 
By contrast, our approach combines the efficiency of online optimization with generally applicable and well-motivated data-processing techniques.
We numerically demonstrate these advantages in several scenarios including mixed states, higher-dimensional systems, and restricted measurements.
\end{abstract}

\maketitle


Quantum information processing (QIP) promises advantages in a wide range of
different contexts, including machine learning
\cite{wiebe_using_2014,wiebe_quantum_2014-2,wiebe_can_2015}, chemistry simulation
\cite{babbush_chemical_2014,hastings_improving_2015,kassal_simulating_2011},
and number theory \cite{shor_polynomialtime_1995,van_dam_quantum_2012}. As such, the
experimental effort to build useful QIP devices has exploded in recent years.
In the course of this effort, \emph{quantum tomography} is a valuable tool for
diagnosing and debugging small quantum devices, and has subsequently seen a
variety of different advances. In particular, Bayesian approaches to tomography 
which are especially well suited to utilizing prior information and adapting to changing experimental conditions 
have developed significantly in recent years
\cite{huszar_adaptive_2012,ferrie_quantum_2014,ferrie_high_2014,granade_practical_2016},
presenting a useful experimental tool
\cite{kravtsov_experimental_2013,struchalin_experimental_2015}.

In this paper we demonstrate the efficiency and accuracy of an adaptive tomography protocol 
that we call PAQT: practical adaptive quantum tomography. PAQT intelligently selects new measurements based on the outcomes of previous ones \cite{fischer_quantum-state_2000,barndorff-nielsen_fisher_2000,rehacek_minimal_2004,bagan_separable_2006,sugiyama_adaptive_2012,huszar_adaptive_2012}. 
Adaptivity has been experimentally demonstrated
\cite{hannemann_self-learning_2002,mahler_adaptive_2013,chapman_experimental_2016,kravtsov_experimental_2013,struchalin_experimental_2015,qi_recursively_2015}, but is not currently standard practice. Though adaptivity increases accuracy, the computational costs incurred outweigh that of simply repeating standard measurements many times. The PAQT approach employs a simple heuristic that can be efficiently computed between measurements, even with embedded hardware \cite{stockton_programmable_2002,casagrande_design_2014,lamb_fpgabased_2016,hornibrook_cryogenic_2015}. 
The algorithm we propose is therefore 
compatible with modern experimental design and avoids an important limitation of previous approaches.

We base our algorithm off of self-guided quantum tomography (SGQT), which treats adaptive tomography as a direct optimization problem rather than a new optimization problem between each measurement
\cite{ferrie_self-guided_2014}. Though this affords an efficient and easy to implement adaptive heuristic, SGQT is not without its limitations. It requires assuming that
the target state is pure, and it does not return rich region estimates
for a state. What PAQT achieves is to effectively combine SGQT with
conventional and easily-implemented tomographic estimators, such as the Bayesian particle filter or
least-squares fit estimators. Under this approach, an experimentalist
can collect data using SGQT (even if its assumptions are not met), and then post-process this data using particle filtering or least-squares fit.

The benefit of PAQT is two-fold. (1) From the point of view of traditional tomography, it gives an adaptive tomography protocol requiring only modest computational resources, as
the bulk of the computational cost is offloaded to post-processing. (2) From the point of view of simulation-based optimization tomography (such as SGQT), it  
effectively augments the output with region estimation providing a statistically robust quantification of uncertainty.

The outline of the paper is as follows. In \autoref{sec:tomographic-problem}
we define and review the problem of tomography as well as three standard
solutions: least squares, maximum likelihood and Bayesian mean estimation. In
\autoref{sec:self-guided}, we review the approaches to measurement adaptive
tomography including the recently introduced self-guided technique. In
\autoref{sec:hybridized}, introduce PAQT by combining SGQT with adaptive Bayesian tomography
and detail the results of our numerical experiments. \autoref{sec:discussion}
concludes with a discussion. 

\section{The Tomographic Problem}
\label{sec:tomographic-problem}

In quantum state tomography, we are interested in reconstructing a quantum
state from a collection of \emph{informationally complete} measurements made
on that state
\cite{newton_measurability_1968,band_empirical_1970,band_quantum_1979}. That
is, a set of measurements is chosen such that if one learns their frequencies
given a quantum system of interest, the frequencies for any other measurement
of that system can then be predicted. If the system of interest is a qubit,
for instance, then knowing the expectations of the observables $\{\sigma_x,
\sigma_y, \sigma_z\}$ allows for predicting the distribution over outcomes for
any other measurement. The empirical reconstruction of quantum states from
measurements of informationally complete observables has been reviewed
by \citet{dariano_quantum_2003}, and reviewed in the case of continuous variables
by \citet{lvovsky_continuous-variable_2009}. Here, we will focus on the case of
state tomography in finite-dimensional systems.

That a quantum state can be empirically determined in principle, however,
leaves the question of how to estimate a state in practice, given finite
experimental resources. For instance, given data from an informationally complete
set of observables, one could use a linear reconstruction,
a maximum likelihood estimator \cite{hradil_quantumstate_1997,christandl_reliable_2012,faist_practical_2015}, or a Bayesian mean estimator \cite{blume-kohout_optimal_2010,huszar_adaptive_2012,ferrie_quantum_2014,ferrie_high_2014,granade_practical_2016} to report a state. 
We will detail each such approach below,
and describe their relative strengths and weaknesses.

Before proceeding, we note that though we consider the general
case of tomography in this work, substantial progress has been made
by considering considering important special cases
under which a state can be much more easily characterized.
In particular, permutationally invariant tomography reconstructs the part of a multiqubit density matrix which is invariant under exchange of the qubits \cite{toth_permutationally_2010}. Compressed
sensing allows for the efficient recovery of \emph{low-rank} quantum states
\cite{gross_quantum_2010,flammia_quantum_2012},
and has been applied experimentally in systems as large as six qubits
\cite{schwemmer_experimental_2014}.
Similarly, MPS \cite{cramer_efficient_2010} and PEPS \cite{landon-cardinal_practical_2012} tomography
use the MPS and PEPS ansatzes to improve exponentially on na\"ive methods for states that are
well-approximated by common tensor network ansatzes \cite{bridgeman_hand-waving_2016}.
Though we do not explore the possibility in this work, we expect that heuristic
approaches should also offer similar advantages to tomographic estimation
in these cases.

\subsection{Problem set up}

First, consider an orthonormal basis for traceless Hermitian operators $\{B_j\}_{j=1}^{d^2-1}$---the Pauli basis, for example. 
That is, for all $i,j$, $B_j^\dag = B_j$ and $\Tr(B_k B_j) = \delta_{kj}$ and $\Tr(B_j) = 0$.  Then, any state $\rho$ can be written
\begin{equation}
    \rho = \frac{\id}{d}+\sum_{j=1}^{d^2-1} \theta_j B_j,
\end{equation}
for some vector of parameters $(\vec\theta)_j=\theta_j$. 
Importantly, these parameters are constrained since $\rho\geq0$.
This poses a problem for many approaches, but there are well-motivated methods which 
produce a valid quantum state starting from a non-physical matrix~\cite{smolin_efficient_2012}.

Let us assume two-outcome test measurements are made such that each measurement outcome is either 1 or 0 and represented by the pair $\{P_k,\id-P_k\}$. The Born rule dictates that the probability to get 1, say, is $\Pr(1|\rho,P_k)=\Tr(\rho P_k)$.  Since the operators $\{B_j\}$ form a basis, we can write
\begin{equation}
    P_k = \frac{\id}{d}+\sum_{j=1}^{d^2-1} p_{kj} B_j,
\end{equation}
and the Born rule vectorizes to
\begin{equation}
    \Pr(1|\rho,P_k)= \Tr(\rho P_k) = \frac{1}{d}+\vec{p}_k^{\rm T} \vec{\theta},
\end{equation}
where $(\vec{p}_k)_j = p_{kj}$. Denote $f_k = \Pr(1|\rho,P_k)$ and $(\vec{f})_k = f_k$.  Also define the matrix $\vec X$ with entries $(\vec X)_{kj}= p_{kj}$. Then the above condenses to
\begin{equation}\label{perfect linear system}
    \vec{f} = \frac{1}{d} + \vec{X} \vec{\theta}.
\end{equation}

If we perform at least $d^2$ such measurements such that the set $\{P_k\}_{k=1}^{d^2}$ is linearly independent, then the probabilities $\vec f$ are sufficient to determine $\rho$ uniquely.  That is, the linear system in \autoref{perfect linear system} has a solution set with a single valid quantum state.  In practice we do not have access to $\vec f$, but only samples drawn from the distribution that
it defines.  Suppose $N_k$ measurements of $\{P_k,\id-P_k\}$ yielded $n_k$ 1s and $N_k-n_k$ 0s.  Then, the empirical frequencies are
\begin{equation}
    \hat{\vec{f}}_k =\frac{n_k}{N_k}.
\end{equation}
The task of tomography is to assign a quantum state $\vec \theta$ to each data set $\hat{\vec f}$.

\subsection{Linear inversion tomography}
\label{sec:lsf}

Next, we will outline the traditional approach to solving the tomography problem.
While we do not recommend this approach, it usually provides reasonable answers and is at least implicitly the starting point for more sophisticated approaches. 

We begin by setting the empirical frequencies equal to the (rescaled) theoretical probabilities $\hat{\vec f} = \vec f$. 
After all, $\Tr(\rho P_k)$ is literally the expectation value of the observable $P_k$. 
In any case, if we let $\vec Y = \hat{\vec f} - 1/d$, the new system of equations
\begin{equation}\label{estimated linear system}
    \vec{Y} = \vec{X} \vec{\theta},
\end{equation}
may not have a solution if more than $d^2$ different measurements have been made.
The traditional approach is to use the \emph{least squares} estimator
\begin{equation}
    \hat{\vec{\theta}}_{\rm LS} = \underset{\vec{\theta}}{\operatorname{argmin}}\|\vec{Y} - \vec{X} \vec{\theta}\|_2^2,
\end{equation}
which has the exact solution
\begin{equation}
    \hat{\vec{\theta}}_{\rm LS} = \left(\vec{X}^{\rm T}\vec{X}\right)^{-1}\vec{X}^{\rm T} \vec{Y}.
\end{equation}
This solution is not guaranteed to produce a positive semidefinite estimate. 
One can resort to performing \emph{constrained} least squares (which is ``not that hard'' since one probably has access to a black box implementation of this using a canned scientific software library) or one can use a two-step approach~\cite{smolin_efficient_2012} that outputs the ``closest'' physical state to a given matrix. 
There is no consensus on which should be preferred and we make no recommendations here.
In our simulations, we have set all negative eigenvalues to zero, as we observe that in practice, measurements designed by self-guided tomography tends to only rarely yield
$\hat{\vec{\theta}}_\LS$ corresponding to $\rho \not\ge 0$.

\subsection{Maximum likelihood tomography}

The linear least squares approach is folklore as old as the problem of tomography, but has been stated explicitly by \citet{qi_quantum_2013}. 
It usually arises when using a Gaussian approximation to the likelihood function in maximum likelihood estimation (see, for example, \citet{kaznady_numerical_2009}). 
The \emph{likelihood function} is the probability distribution of the data \emph{given} a state $\vec\theta$, thought of as a function of $\vec\theta$. 
Since each measurement is an independent binomial trial, the likelihood function is quite simple:
\begin{widetext}
\begin{equation}
    \label{eq:likelihood}
   \Pr(\hat{\vec f}|\vec\theta, \vec X) = \prod_{k} \binom{N_k}{N_k\hat f_k}\left( \frac1 d + \vec{p}_k^{\rm T} \vec{\theta} \right)^{N_k \hat f_k}\left(1- \frac1d- \vec{p}_k^{\rm T} \vec{\theta} \right)^{N_k (1-\hat f_k)}.
\end{equation}
\end{widetext}
One of the oldest techniques in classical statistical estimation is \emph{maximum likelihood estimation} (MLE), which prescribes the estimate 
\begin{equation}
    \hat{\vec{\theta}}_{\rm MLE} = \underset{\vec{\theta}}{\operatorname{argmax}}\Pr(\hat{\vec f}|\vec\theta, \vec X).
\end{equation}
This does not have a closed form in general. To make some traction, we can approximate the likelihood function by a Gaussian (perhaps with appeal to the central limit theorem).  A Gaussian is defined by its mean and variance, so we need only those from the actual distribution to make the approximation.  These are simple enough to derive from the properties of the binomial distribution:
\begin{align}
\mathbb E[\hat{\vec f}] & = \frac1d + \vec X \vec{\theta},\\
\mathbb V[\hat{\vec f}]_{kj} & = \delta_{kj} \frac{\left( \frac1 d + \vec{p}_k^{\rm T} \vec{\theta} \right)\left(1- \frac1d- \vec{p}_k^{\rm T} \vec{\theta} \right)}{N_k}.
\end{align}

The location of the maximum of a function is the same as that of the log of the function.  The logarithm of the Gaussian approximation to the likelihood function (ignoring terms which do not depend on $\vec\theta$) is
\begin{equation}
    -\frac12 \sum_k \frac{(Y_k -\vec{p}_k^{\rm T} \vec{\theta})^2  N_k }{\left( \frac1 d + \vec{p}_k^{\rm T} \vec{\theta} \right)\left(1- \frac1d- \vec{p}_k^{\rm T} \vec{\theta} \right)}. 
\end{equation}
We make one more approximation, which is again replacing the probabilities with their empirical frequencies \footnote{A discussion of the consistency of this replacement can be found in \cite{berkson_estimation_1956}.
} such that the maximum likelihood problem then becomes
\begin{equation}
    \hat{\vec{\theta}}_{\rm MLE} = \underset{\vec{\theta}}{\operatorname{argmin}}\|\vec{Y}' - \vec{X}' \vec{\theta}\|_2^2,
\end{equation}
where we have \emph{weighted} $\vec Y$ and $\vec X$ by the variance:
\begin{equation}
Y_k'  = \sqrt\frac{N_k}{\hat f_k (1-\hat f_k)} Y_k, \; 
X_k'  = \sqrt\frac{N_k}{\hat f_k (1-\hat f_k)} X_k.
\end{equation}
Notably, this approach fails if $\hat{f}_k = 0$ or $1$ for any $k$,
as the variance in these cases approaches zero, so that $Y'_k \to \infty$.
To solve this, we hedge the empirical frequencies by $\beta = 0.5$, so that
we use $\hat{f}_k = (n_k + 0.5) / (N_k + 1)$ when computing the MLE
\cite{blume-kohout_hedged_2010}.

\subsection{Bayesian tomography}
\label{sec:bayes}

As opposed to the frequentist techniques noted above, the Bayesian approach centers around Bayes' rule,
which prescribes how to update a prior distribution $\Pr(\vec{\theta})$ to a posterior distribution $\Pr(\vec{\theta} | \hat{\vec{f}}, \vec{X})$
that is conditioned on the observed frequencies $\hat{\vec{f}}$. Concretely,
\begin{equation}
    \label{eq:posterior}
    \Pr(\vec\theta|\hat{\vec f}, \vec X)= \frac{\Pr(\hat{\vec f}|\vec\theta, \vec X) \Pr(\vec\theta)}{ \Pr(\hat{\vec f}| \vec X)},
\end{equation}
where $\Pr(\hat{\vec f}|\vec\theta, \vec X)$ is the likelihood function of \autoref{eq:likelihood},
and where $\Pr(\hat{\vec f}| \vec X)$ is a pesky normalization that we will deal with implicitly when
doing numerical calculation.
When Bayes' rule is used iteratively, the posterior for one experiment becomes the prior for the next.
In words, this equation is a prescription of the full distribution of knowledge about the quantum state \emph{given} the data that was actually observed. What can we do with this? 

First, we can produce a single ``point'' estimate of $\vec\theta$ via the posterior mean:
\begin{equation}
    \label{eq:bme-point-estimate}
    \hat{\vec\theta}_{\rm BME} = \mathbb E_{\vec\theta|\hat{\vec f}, \vec X}[\vec\theta],
\end{equation}
where BME stands for Bayesian mean estimator. 
The mean estimator is not the only option, though it is optimal for certain figures of merit \cite{blume-kohout_optimal_2010}, or at least near-optimal \cite{kueng_near-optimal_2015}. 
Second, the posterior distribution naturally encodes ``error bars'' by way of the posterior covariance tensor \cite{blume-kohout_optimal_2010,granade_practical_2016}. 
Finally, the data can be processed \emph{online} in the sense that new data can be incorporated into the distribution without the need to reanalyze all previous data at the same time. 
This lends itself naturally to \emph{adaptive} tomography, discussed in the next section.

In practice, however, exactly implementing Bayesian mean estimation is quite
difficult, as the expectation value in \autoref{eq:bme-point-estimate} may not
be analytically tractable outside of important special cases. 
We will therefore
follow the approach of \citet{huszar_adaptive_2012} and use the particle filtering
algorithm \cite{doucet_tutorial_2011} to numerically implement Bayesian estimation.
This approach has since been used by \citet{ferrie_quantum_2014,ferrie_high_2014}
and by \citet{granade_practical_2016} to develop useful applications of Bayesian
tomography, by \citet{stenberg_adaptive_2015} to learn coherent states,
and has been successfully applied outside of tomography to
efficiently learn Hamiltonians using classical \cite{granade_robust_2012}
and quantum resources \cite{wiebe_hamiltonian_2014}.

Particle filtering proceeds by approximating the prior and posterior distributions
at each step of Bayesian inference as a weighted sum of $\delta$ functions,
\begin{equation}
    \label{eq:smc-approx}
    \Pr(\vec{\theta}) \approx \sum_i w_i \delta(\vec{\theta} - \vec{\theta}_i),
\end{equation}
where $\{w_i\}$ are the weights of the \emph{particles} located at
$\{\vec{\theta}_i\}$. Upon observing a datum $\hat{f}_k$, the weights are then updated
by calling the likelihood function for each particle,
\begin{equation}
    \label{eq:smc-update}
    w_i \mapsto w_i \times \Pr(\hat{f}_k | \vec{\theta}_i) / \mathcal{N},
\end{equation}
where $\mathcal{N}$ is the normalization factor in Bayes' rule \autoref{eq:posterior},
which can be found implicitly
by demanding that $\sum_i w_i = 1$. The BME is then found by taking a sum over
the particles representing the current posterior,
\begin{equation}
    \label{eq:smc-bme}
    \hat{\vec{\theta}}_{\mathrm{BME}} = \sum_i w_i \vec{\theta}_i.
\end{equation}
Numerical stability in particle filtering is provided by the use of a
\emph{resampling algorithm} which replaces the particles by a new set of particles
that more effectively represents the same posterior. We will use the Liu and West
resampling algorithm \cite{liu_combined_2001}, which mixes the current posterior
with a Gaussian distribution of the same mean and covariance. The resampling
is controlled by a parameter $a \in [0, 1]$, with smaller $a$ corresponding
to ``more Gaussian'' posteriors. 

\section{Adaptive and Self-Guided Tomography}
\label{sec:self-guided}

We have not yet addressed the issue of $\vec X$, the matrix defining the choice of measurements. 
How should this choice of measurements be done? 
This is an open problem, with the lack of consensus mostly due
to incompatible choices of criteria for optimality.  In any case,
the fact that some measurements are better than others suggests that
improvements can be made through \emph{adaptive} tomography---that is, choosing
new measurement settings based on information obtained from past measurement
settings.

\subsection{Adaptive tomography}

The first to consider adaptive state tomography was \citet{fischer_quantum-state_2000}, who did so for a single qubit assumed to be in a pure state. 
That is, the prior was taken to be a uniform distribution on the surface of the Bloch sphere. 
The adaptivity consists of maximizing the entropy of the sampling distribution and expected fidelity. 
The estimator was chosen to be the maximum of the posterior distribution. 
This was later experimentally realized for a short set of measurements by pre-computing and storing the optimal experiment choices in a look-up table \cite{hannemann_self-learning_2002}. 

Adaptive state tomography has also been investigated in the context of parameterized models and Fisher information. 
\citet{barndorff-nielsen_fisher_2000} showed that the quantum Fisher information for a single parameter can be obtained asymptotically by adaptively choosing the measurement settings in a two-stage procedure. 
The asymptotic two-step approach seems also to have been independently discovered by  \citet{rehacek_minimal_2004} and \citet{bagan_separable_2006}.
An experimental demonstration has verified a quadratic improvement in accuracy \cite{mahler_adaptive_2013,hou_achieving_2016}. 
These approaches, however, are of more theoretical interest as they are guaranteed only asymptotically or require the total number of measurements to be specified \emph{a priori}.

A generic approach using the maximum likelihood estimator and measurements minimizing the expected variance also showed an improvement over standard quantum tomography \cite{sugiyama_adaptive_2012}. 
This has been made more practical through use of a recursive least-squares formula in \citet{qi_recursively_2015}. Below we will see that our choice of heuristic for adaptation may lead the least squares estimator to fail due to ill-conditionedness.
Our results below will suggest that the better approach is the Bayesian one.

\subsection{Bayesian adaptive quantum tomography}

The Bayesian method also allows for a principled approach to adaptive
measurements since one has a very formal definition of \emph{expected utility} of a
measurement. 
Consider \autoref{eq:posterior} in the case of a \emph{hypothetical} measurement $\vec X$, which could produce data $\hat{\vec f}$. 
Then, one can define the expected utility of the measurement as
\begin{equation}
    U(\vec X) = \mathbb E_{\hat{\vec f}, \vec \theta|\vec X}[L(\vec \theta)],
\end{equation}
where $L$ is an arbitrary loss function. 

\citet{fischer_quantum-state_2000} considered both the log-loss and fidelity for a single qubit. \citet{huszar_adaptive_2012} considered the \emph{information gain}, which has since been
used to define an adaptive protocol in one- and two-qubit optical experiments \cite{kravtsov_experimental_2013,struchalin_experimental_2015}. 
Most recently, the fidelity for arbitrary dimensions has been studied and numerics performed on one and two-qubits \cite{kalev_fidelity-optimized_2015}.

Calculating these utilities, however, poses a problem since one
may be able to perform a great deal of non-optimized experiments before the calculation of the ``best'' experiment can be completed. 
These intermediate experiments, while not optimized, still contain useful information about the state and may provide better accuracy when the cost of optimization is included. 
Hence the need for \emph{heuristics} that realize the benefits of adaptivity while avoiding
the costs of explicit optimization over utility functions. 

In the context of Hamiltonian learning,
for example, heuristics have been used to obtain many of the benefits of explicitly
optimizing a utility, while avoiding much of the computational expense 
\cite{ferrie_how_2013,wiebe_hamiltonian_2014}. 
Machine learning techniques
have recently been applied to the design of good heuristics for quantum characterization
problems \cite{stenberg_characterization_2015}, but we will take a different
approach and instead use \emph{stochastic optimization}
to provide an efficient heuristic.

\subsection{Self-guided quantum tomography}

Self-guided quantum tomography (SGQT) is an adaptive tomography scheme which
avoids the linear inversion problem altogether by posing the tomography
problem as one of optimization rather than estimation
\cite{ferrie_self-guided_2014}. In particular, self-guided tomography
finds a pure state $\ket{\phi}$ such that the overlap $F(\phi , \rho)
= \braket{\phi | \rho | \phi}$ is maximized for a true state $\rho$.
If $\rho = \ket{\psi}\!\!\bra{\psi}$ is a pure state, then $F(\phi , \rho)$
is maximized if and only if $\ket{\phi} = \e^{\ii \theta} \ket{\psi}$
for a phase $\theta$, such that an optimal solution is also an accurate
estimate of the true state.

An earlier work took a similar approach by testing whether the unknown qubit
state was symmetric with a reference state \cite{happ_adaptive_2008}, where the
reference state is chosen adaptively to maximize fidelity. 
However, the method
is defined only for a single qubit and requires a second fully characterized and
controllable qubit along with an entangled measurement.

Having phrased state estimation as an optimization problem, self-guided
tomography proceeds by experimentally estimating the objective function $F$
from empirical frequencies. This results in a stochastically evaluated
objective function, such that the optimization problem is amenable to attack
by stochastic optimization algorithms. We will in particular rely on the
simultaneous perturbative stochastic algorithm
(SPSA) \cite{spall_multivariate_1992}.

The SGQT estimate is precisely defined as follows. We begin with a random state $\ket{\phi_0}$ and iteratively produce new states $\ket{\phi_{k}}$ which serve the dual role of specifying the current estimate of the state and next measurements to perform. 
At iteration $k$, we perform the measurements $\{P_{k,\pm}, \id - P_{k,\pm}\}$, where 
\begin{equation}\label{eq:SGQT_meas}
P_{k,\pm} = \ket{\phi_{k-1} \pm\epsilon_{k} \Delta_{k}}\!\bra{\phi_{k-1} \pm\epsilon_{k} \Delta_{k}},
\end{equation}
and $\Delta_{k}$ is a random vector that is constructed by setting each entry to $\pm 1$ with equal probability. 
Here $\epsilon_{k}$ is a step-size parameter chosen below.
The outcomes of these measurements are denoted $\hat f_{k,\pm}$. The gradient of the fidelity is estimated from these measurements to be
\begin{equation}
\hat g_{k} = \frac{\hat f_{k,+}-\hat f_{k,-}}{2\epsilon_{k}} \Delta_{k}.
\end{equation}
Using these, and an additional gain parameter $\alpha_{k}$, the SPSA algorithm mimics standard gradient ascent, but along the random direction $\Delta_{k}$:
\begin{equation}
\ket{\phi_{k}} = \ket{\phi_{k-1} +\alpha_{k} \hat g_{k}}.
\end{equation}
Convergence is guaranteed \cite{spall_multivariate_1992} given the specification of $\Delta_k$ above and 
\begin{subequations}
\begin{align}
\epsilon_{k} &= \frac{1}{k^{1/3}},\\
\alpha_k &=  \frac1k.
\end{align}
\end{subequations}
Unless otherwise noted, however, we shall use the parameters suggested
by \citet{spall_multivariate_1992},
\begin{align}
    \epsilon_{k} = 0.1 / k^{0.101} \text{ and } \alpha_k = 10 / k^{0.602}.
\end{align}

SPSA has also been applied in
quantum information to design high-fidelity control sequences given randomized
benchmarking experiments \cite{ferrie_robust_2015,granade_characterization_2015}. 
In particular, Ferrie showed that self-guided tomography can rapidly learn pure states for comparatively large quantum systems \cite{ferrie_self-guided_2014}.
To the best of our knowledge, self-guided tomography is the only adaptive tomography technique which has gone beyond two qubits, even in simulation.
SGQT has also recently been demonstrated in an optical experiment \cite{chapman_experimental_2016}.

SQGT is not without its limitations, however. The aim of the current work is
to mitigate the following three limitations of SGQT: (1) it is restricted to
pure state tomography, (2) it does not report error bars, and (3) it can not
be restricted to local measurements.

\section{Practical Adaptive Quantum Tomography}
\label{sec:hybridized}

\begin{figure}[t!]
    \begin{centering}
        \includegraphics[width=\columnwidth]{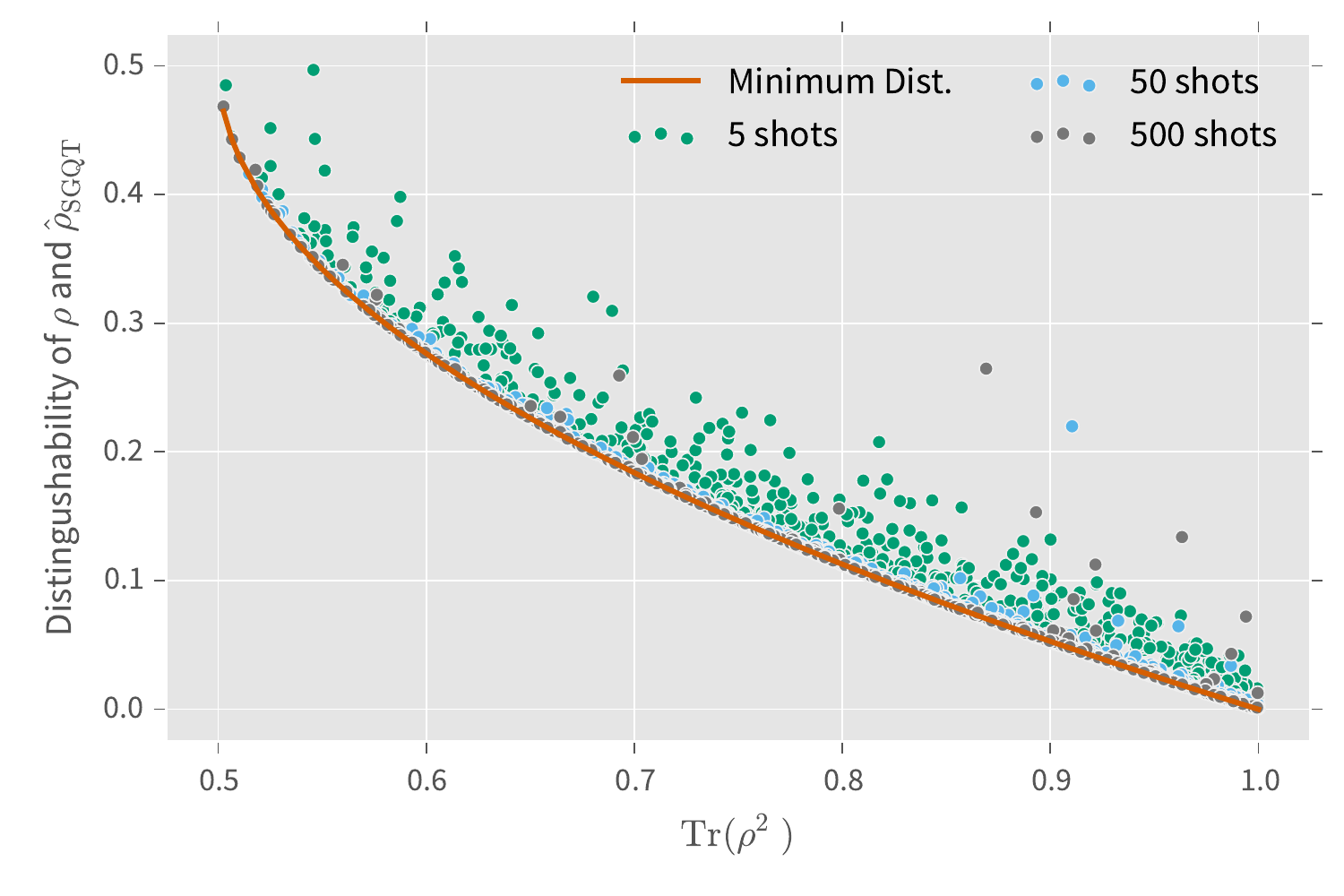}
    \end{centering}
    \caption{
        \label{fig:min-distinguishability}
        Distinguishability between self-guided estimated states and true states 
        drawn from the Hilbert-Schmidt prior for a qubit, plotted versus the best
        achievable distinguishability for any estimator constrained to pure states,
        where he distinguishability between $\rho$ and $\sigma$ is defined as the trace distance $\frac12\|\rho-\sigma\|_1$.
        The self-guided estimates are drawn from 10,000 iterations with either 5,
        50 or 500 shots per measurement. As the number of shots per measurement
        increases, the self-guided estimates approach the closest states allowed
        by the pure state assumption, demonstrating that the self-guided procedure
        produces useful data even when the true state is mixed.
    }
\end{figure}

From the above discussion, we find that self-guided quantum
tomography potentially offers many advantages for experimental practicality over
traditional protocols, but at the cost that it does not accurately report
mixed states, and does not certify its own errors. 
Happily, these are
precisely the advantages of the Bayesian approach, such that we can collect
data using self-guided tomography, then post-process with offline estimation. 

We introduce PAQT---practical adaptive quantum tomography---an optimized numerical algorithm which implements the idea of merging self-guided tomography as an online experiment design heuristic into Bayesian data analysis. 
Our algorithm automatically selects experiments online and can be implemented with modest experimental hardware, including modern embedded controllers such as field-programmable gate arrays (FPGAs). 
The advantages of PAQT are that it provides the enhanced precision of adaptive tomography together with fast data processing and experiment design. 
The framework provides robust and easily interpretable error regions without additional overhead. 
Explicitly, PAQT uses the results of the measurements \autoref{eq:SGQT_meas} specified by SGQT with the Bayesian mean estimator \autoref{eq:bme-point-estimate}.
Although we demonstrate the algorithm for state tomography, the method is equally applicable to channel tomography and other estimation tasks and can easily accommodate other estimators. 

\begin{figure*}
    \begin{centering}
        \includegraphics[width=\textwidth]{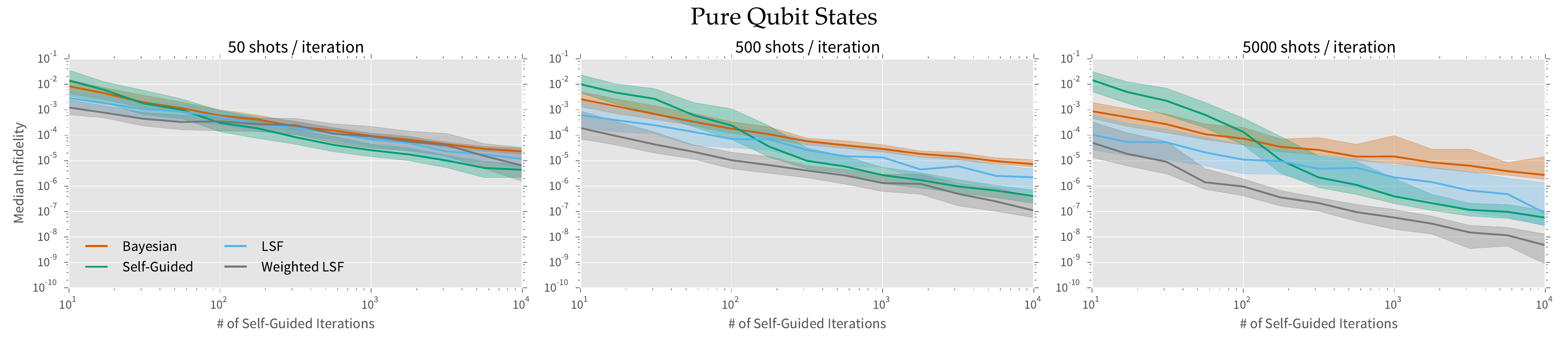}

        \includegraphics[width=\textwidth]{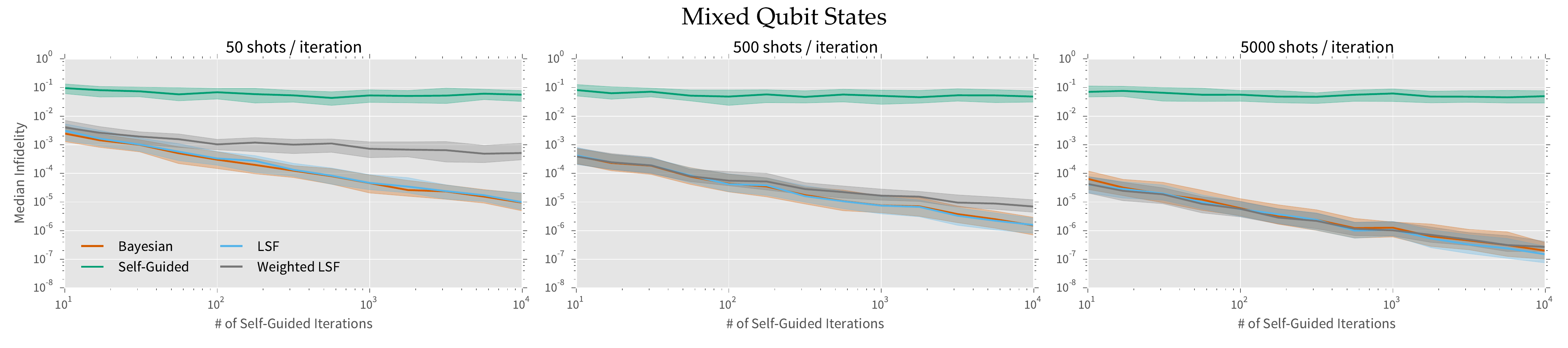}
    \end{centering}
    \caption{
        \label{fig:qubit-infid}
        Median infidelity $r = 1 - F$ for self-guided tomography on single qubit (top) pure
        and (bottom) mixed states, both without post-processing the self-guided data (green),
        as well as post-processing via PAQT using Bayesian (orange) and least-squares estimators (gray and blue). In both cases,
        Bayesian tomography is performed with a full-rank (Hilbert-Schmidt)
        prior, using the particle filter summarized in \autoref{sec:bayes}
        with 4,000 particles and the resampling parameter $a = 0.98$.
        The shaded regions indicate the 16\%
        and 84\% quantiles over trials. Note that, for a normal distribution,
        this region would coincide with the $1\sigma$--confidence interval,
        but as illustrated in \autoref{fig:qubit-loss-density}, the losses
        are far from normally distributed, such that we cannot make the
        normal interpretation.
        The self-guided procedure works very well for pure states (top),
        providing estimates with fidelity approximately $99.999\%$ after
        $10^{7}$ bits of data. For mixed states, the self-guided procedure
        does not learn well on its own, but post-processing the self-guided
        data with Bayesian or least-squares estimation produces high-fidelity
        estimates.
    }
\end{figure*}

Our results use the QInfer 1.0a1
\cite{granade_qinfer_2012},
QuTiP \cite{johansson_qutip_2013} 3.2.0, NumPy \cite{walt_numpy_2011} 1.9.2, Pandas \cite{pandas}
0.16.2 and SciPy 0.15.1 \cite{jones_scipy:_2001} libraries for Python 2.7 (Enthought Canopy 1.5.4) to perform the Bayesian analysis. We performed
all simulations on the University of Sydney School of Physics cluster. Full source code for our
simulations, and for our implementations of self-guided and least-squares
tomography can be found online \cite{supp_mat}.

We start by noting in \autoref{fig:min-distinguishability} that, in the case
of qubits, the states estimated by self-guided tomography are almost as indistinguishable
as the pure state closest to each true state in terms of the 1-norm. This makes it
clear that, although self-guided tomography should not be expected to return a useful
estimate if the true state is mixed, it is still heavily dependent on the true state
such that we should expect self-guided tomography to collect useful data.

Indeed, as we show in \autoref{fig:qubit-infid}, PAQT effectively combines self-guided tomography with least-squares and Bayesian estimators for both pure and mixed states on a qubit. 
In particular, even though self-guided tomography has ceased to learn states
when the true state is a mixed state, the data collected can be used by both the
Bayesian and least-squares fit (LSF) tomographic estimators to return very good estimates of the state.

\begin{figure*}
    \begin{centering}
        \includegraphics[width=\textwidth]{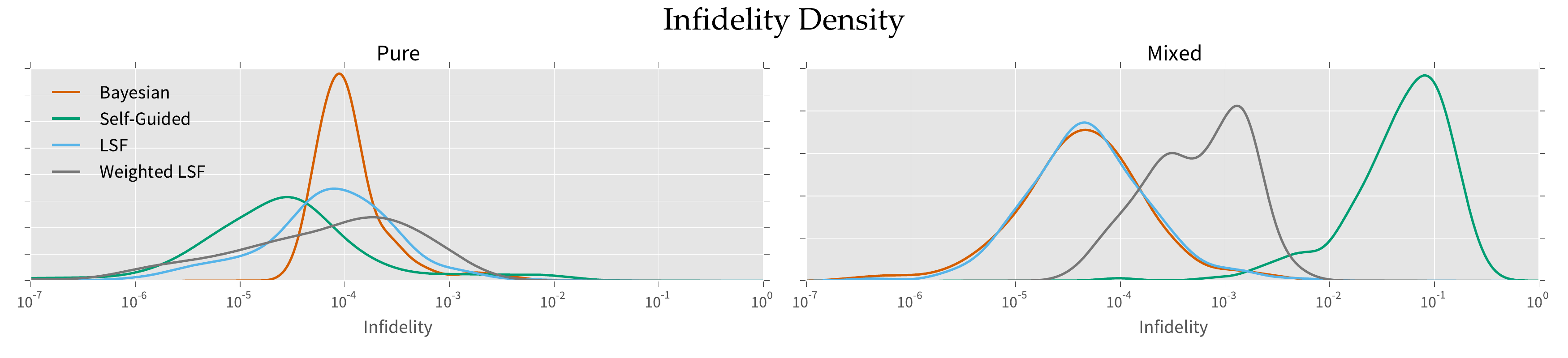}

        \includegraphics[width=\textwidth]{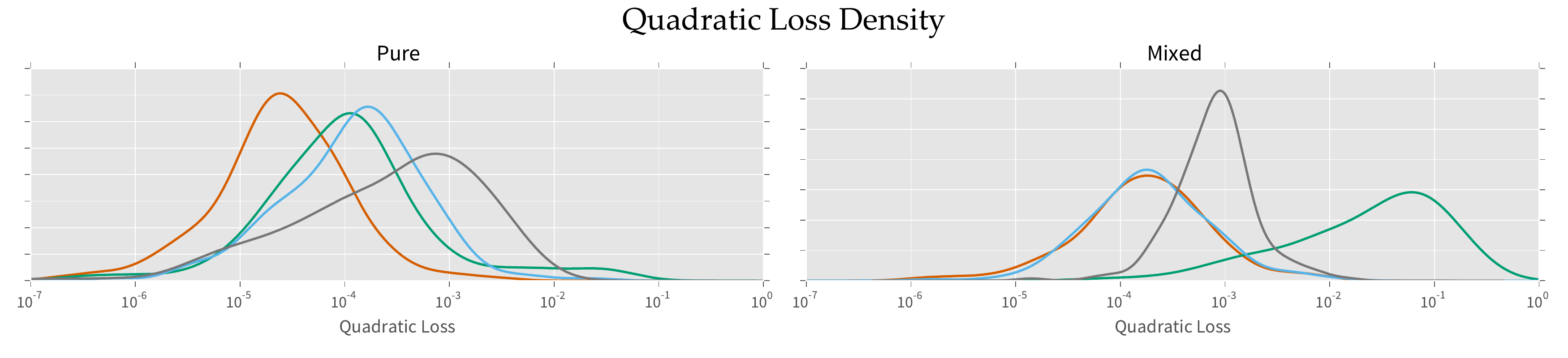}
    \end{centering}
    \caption{
        \label{fig:qubit-loss-density}
        Kernel density estimate of the distribution over losses
        for self-guided quantum tomography without post-processing, as well as PAQT which post-processes the SGQT data using Bayesian and least-squares fit estimators. 
        Tomography simulations are shown for single-qubit pure and mixed states. 
        The top shows the density
        over the infidelity, as directly optimized by self-guided tomography,
        while the bottom shows the density over the quadratic loss. When
        measuring the performance of each algorithm using the infidelity,
        self-guided tomography is optimal for pure states, while Bayesian
        and least-squares post-processing provide the best estimates for
        mixed states. On the other hand, if we use the quadratic loss to
        characterize estimation performance, Bayesian post-processing produces
        the best estimates even in the pure-state case.
    }
\end{figure*}

Which estimator in particular gives the lowest error depends strongly, however,
on the loss function that one uses to quantify error. In \autoref{fig:qubit-loss-density},
we compare the distribution over losses for the four tomographic procedures as applied
to qubit pure and mixed states, and as measured by the infidelity and quadratic loss
functions. Whereas self-guided tomography directly optimizes the infidelity, we note
that it performs very well according to this measure in the pure-state case. Similarly,
the Bayesian mean estimator is optimal for Bregman divergences such as the quadratic
loss $L(\vec{\theta}, \hat{\vec{\theta}}) \defeq (\vec{\theta} - \hat{\vec{\theta}})^\T (\vec{\theta} - \hat{\vec{\theta}})$,
so that it performs very well if we choose to quantify errors accordingly.

\begin{figure*}
    \begin{centering}
        \includegraphics[width=\textwidth]{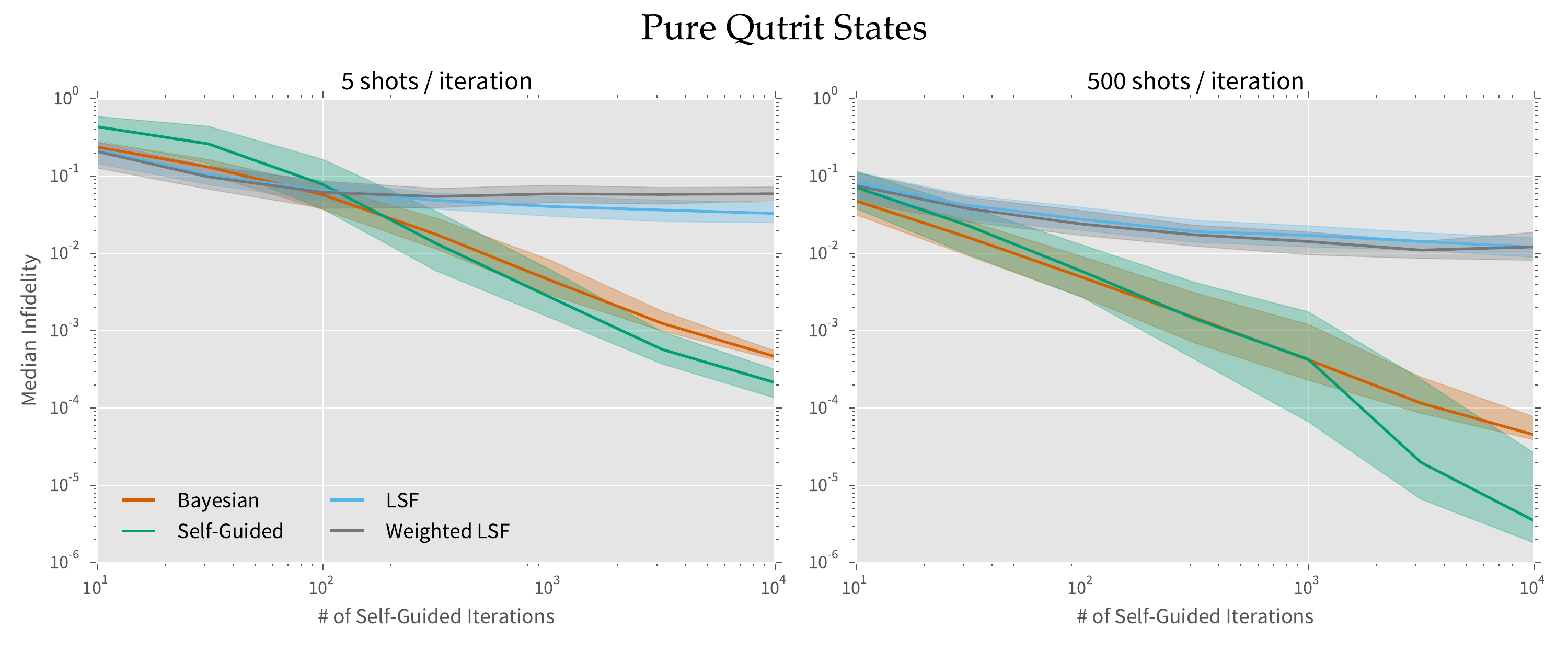}

        \includegraphics[width=\textwidth]{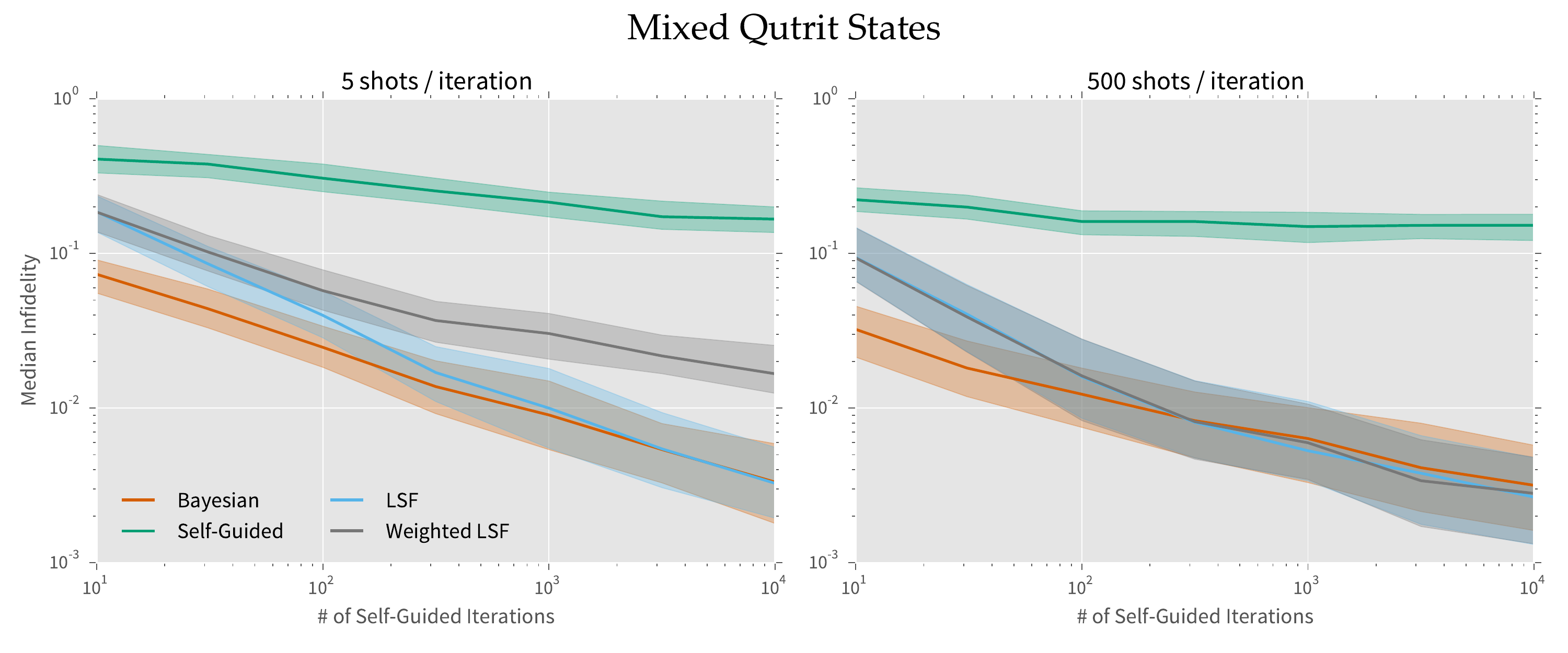}
    \end{centering}
    \caption{
        \label{fig:qutrit-infid}
        Median infidelity for self-guided tomography on single-qutrit (top) pure
        and (bottom) mixed states. In both cases, PAQT is performed with
        Bayesian post-processing using a full-rank (Hilbert-Schmidt)
        prior, 32,000 particles and the resampling parameter $a = 0.9$. The shaded regions indicate the 16\%
        and 84\% quantiles over trials. In this case, Bayesian estimation via PAQT 
        produces high-quality estimates for pure and mixed true states.
        \vspace{2\baselineskip}
    }
\end{figure*}

In \autoref{fig:qutrit-infid}, we consider self-guided tomography of pure and mixed
qutrit states, showing that the benefits of using PAQT to combine SGQT with Bayesian tomography
persist in this case.
Notably, least-squares fitting does significantly less well for self-guided
datasets on pure qutrits. Reducing the resampling parameter $a$ to $0.9$
allows the Bayesian estimator to remain robust in this case, however.

\begin{figure*}
    \begin{centering}
        \includegraphics[width=\textwidth]{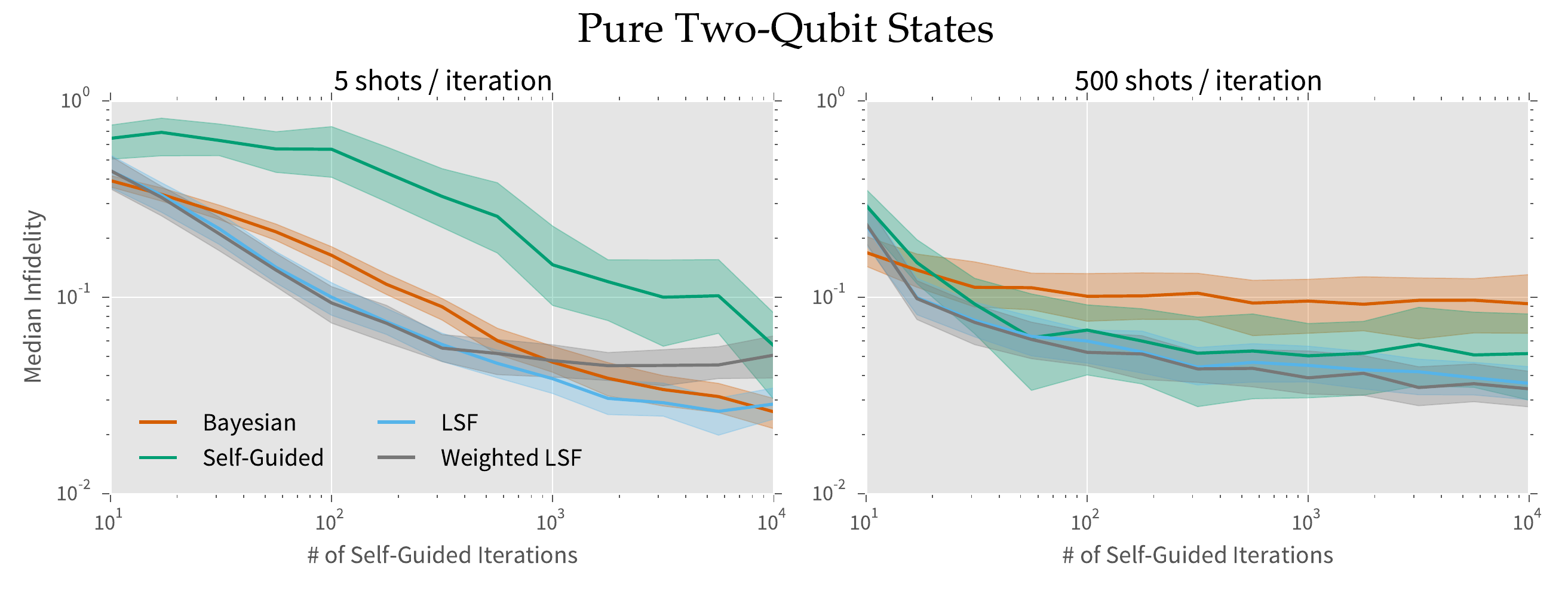}

        \includegraphics[width=\textwidth]{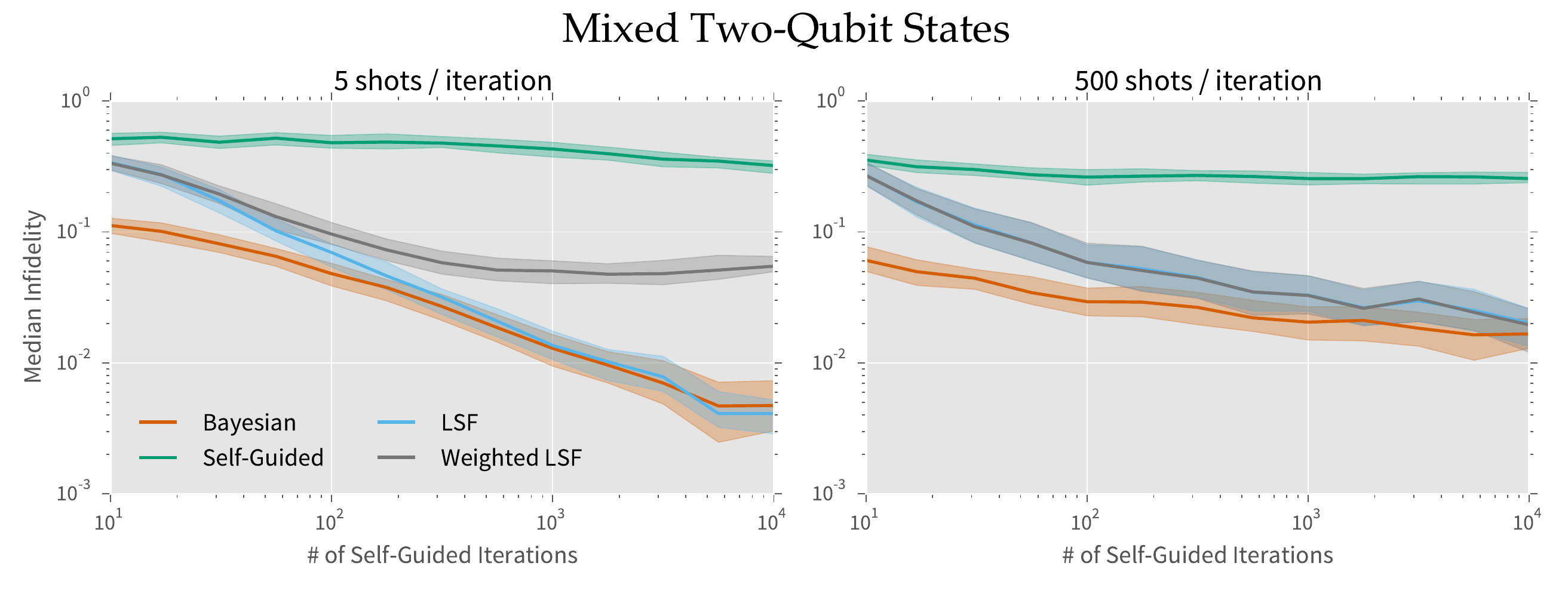}
    \end{centering}
    \caption{
        \label{fig:two-qubit-infid}
        Median infidelity for self-guided tomography on pure (top) and mixed
        (bottom) states of two qubits, restricted to product measurements.
        In both cases, we use PAQT for post-processing with 32,000 particles for the Bayesian estimator,
        and the self-guided tomography data is collected with a gain of $\alpha_k
        = 31 / k^{0.602}$ and a step of $\epsilon_k = 0.1 / k^{0.101}$.
    }
\end{figure*}

We also consider the case in which the optimization procedure used by
self-guided tomography is restricted to an incorrect model of the system under
study. In particular, in \autoref{fig:two-qubit-infid}, we collect data under the
restriction that the true state is a mixed or pure product state of two qubits,
then draw the true state from a Haar or Hilbert-Schmidt prior on the full
four-dimensional state. In this way, the self-guided algorithm is explicitly
following an incorrect model for the state. We note that, despite this, the
Bayesian and least-squares estimators are both able to improve on their initial
uncertainty by using data collected from the product state measurements.

\begin{figure*}
    \begin{centering}
        \includegraphics[width=0.9\textwidth]{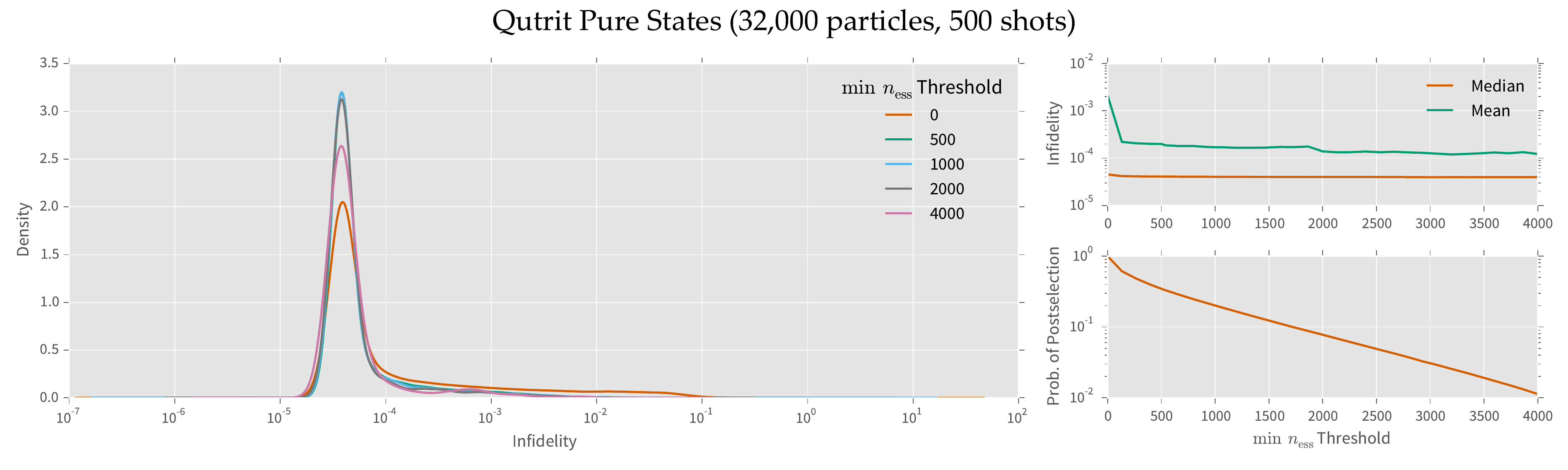}

        \includegraphics[width=0.9\textwidth]{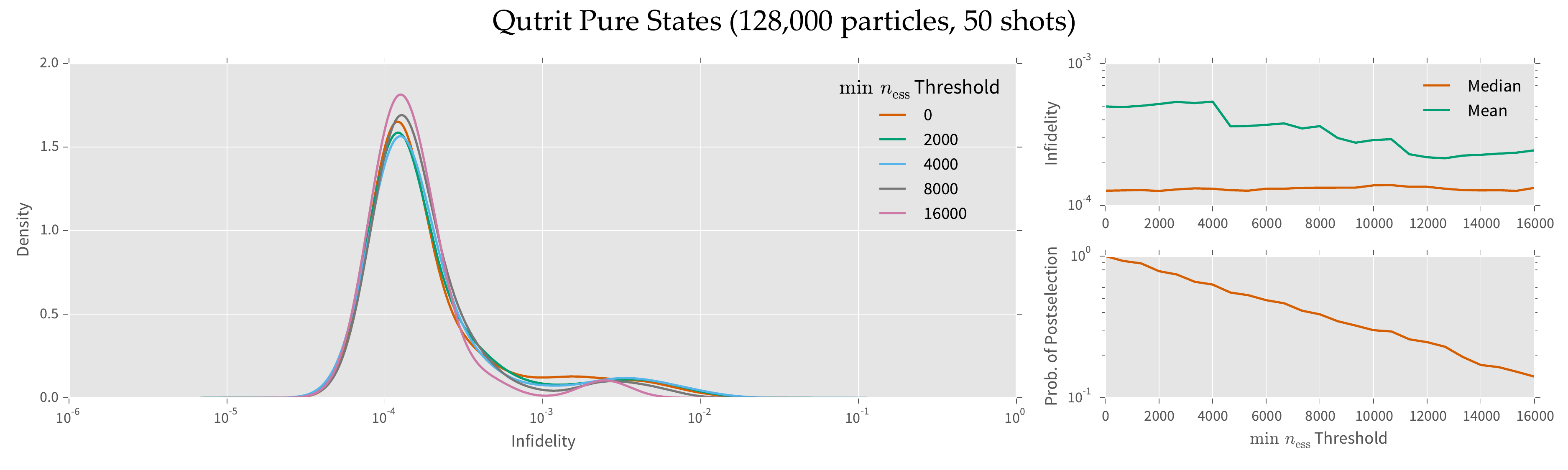}

        \includegraphics[width=0.9\textwidth]{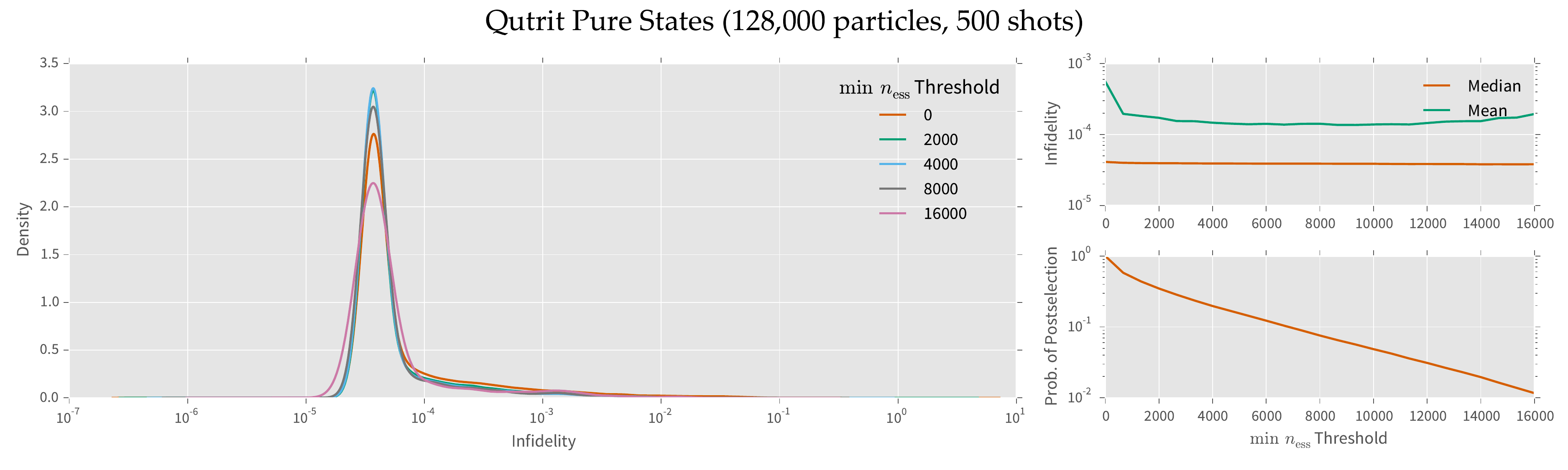}

        \includegraphics[width=0.9\textwidth]{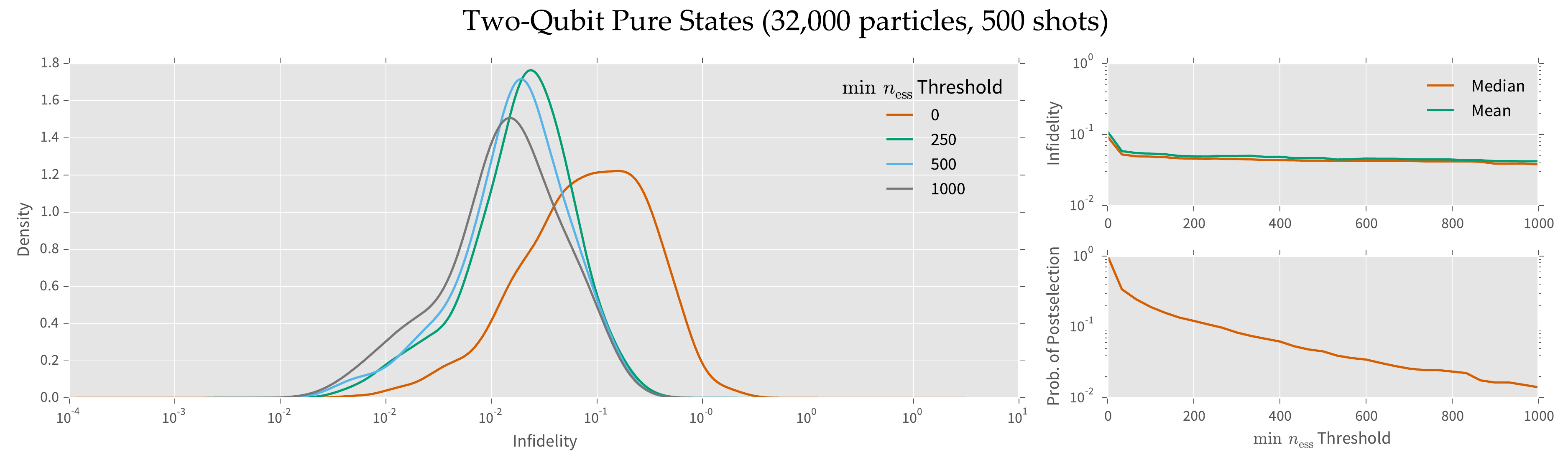}
    \end{centering}
    \caption{
        \label{fig:postsel}
        Performance of PAQT Bayesian post-processing when postselecting on trials
        during which the effective sample sizes $n_{\mathrm{ess}}$ remains
        above various thresholds during out the estimation procedure, for
        qutrit data and for two-qubit data restricted to product measurements.
        For each
        of the three data sets, the left-hand subfigure shows the kernel density
        estimate over infidelity, demonstrating that more demanding thresholds
        can ``shift'' the distribution over infidelity, especially for the
        product-measurement case. The upper-right subfigures for each data set
        show the approach of the mean infidelity to the median fidelity as
        a function of the post-selection threshold, while the lower-right
        subfigures show the probability of the postselection succeeding.
        Importantly, in three of the four cases, we observe that post-selection on
        the diagnostics produced by Bayesian particle filtering can help
        eliminate trials with less accurate estimates. For the case in which
        both a large number of particles are used and a large amount of data
        is taken, the effect of post-selecting on diagnostics is much less
        pronounced.
    }
\end{figure*}

Finally, we note that the performance of the Bayesian estimator
can be dramatically improved if we postselect on diagnostic information provided
by the particle filtering algorithm. 
In \autoref{fig:postsel}, we show the
kernel-density estimated distribution over infidelity for each of the qutrit and
two-qubit cases, postselecting on the smallest effective sample size
observed during a tomography run. 
That is, we accept a tomography trial if the particle filter weights $\{w_i\}$ satisfy
\begin{equation}
    \frac{1}{\sum_i w_i^2} \ge n_{\text{th}}
\end{equation}
throughout the experiment, for some choice of threshold $n_{\text{th}}$.
For the qutrit case, using either 32,000 or 128,000
particles, we observe that as we increase this threshold (that is, as we demand
a larger effective sample size), the mean performance rapidly approaches the
median performance. 
Thus, performing this postselection allows us to exclude the
worst-case performance of the Bayesian estimator. 
On the other hand, when the
data are not especially informative, as in the two-qubit product measurement
case, the benefit of postselection is significantly less pronounced.

\section{Discussion}
\label{sec:discussion}

Though the point of SGQT is to avoid solving a large system of linear equations, the data collected from the performed measurements still define a set of equations that can be inverted in one way or another. This is the approach of LSF and weighted LSF. However, we note that these approaches do not perform well in all but a few of the cases considered. The explanation for this observation is that the constructed linear system is in general ill-conditioned.

Given infinite precision data, SGQT measurements would trace out a straight path through state space from the initial guess to the true state, following the gradient of the fidelity. This set of measurements will not be informationally complete. 
Due to the stochasticity of the algorithm, for finite data, a sufficiently large number of SGQT iterations will be informationally complete, but most of the measurements will be linearly dependent. This frustrates the stability of attempting to solve the linear equations defined by \autoref{estimated linear system}. The standard approach to quantify the stability of a linear system is through the condition number
\begin{equation}
\kappa(\boldsymbol{X}) = \frac{\sigma_1(\boldsymbol{X})}{\sigma_{d^2}(\boldsymbol{X})},
\end{equation}
where $\sigma_1(\boldsymbol{X})$ is the largest and $\sigma_{d^2}(\boldsymbol{X})$ is the smallest singular value. Smaller condition numbers lead to more stable linear systems. We will argue and demonstrate that self-guided tomography leads to measurements which define a linear system with large condition number. 
Importantly, it is only the process by which data is gathered (rather than analyzed) that determines the condition number. 
We will therefore restrict our discussion of condition numbers to SGQT as a data gathering procedure.

The largest singular value will be related to the total number of SGQT iterations since most of the late measurements will be nearly co-linear, clustering around the true state. The smallest singular value would be 1 in the ideal case of performing a subset of orthogonal basis measurements. However, as noted above, the system is only \emph{barely} informationally complete---the matrix $\boldsymbol{X}$ is nearly rank-deficient (rank$<d^2-1$), in other words. The actually value of $\sigma_{d^2}(\boldsymbol{X})$, and hence, $\kappa(\boldsymbol{X})$, will vary quite a bit from run to run, but the scaling with the total number of measurements, $K$, will be $O(\sqrt{K})$. This is because most of the measurements will be approximately co-linear. In the exact case where $\boldsymbol{X}$ consists of $(d^2-1) \times (d^2 -1)$ orthogonal submatrix and $K-1$ repeated rows, the condition number is identically $\sqrt{K}$.

\begin{figure}
    \begin{centering}
        \includegraphics[width=\columnwidth]{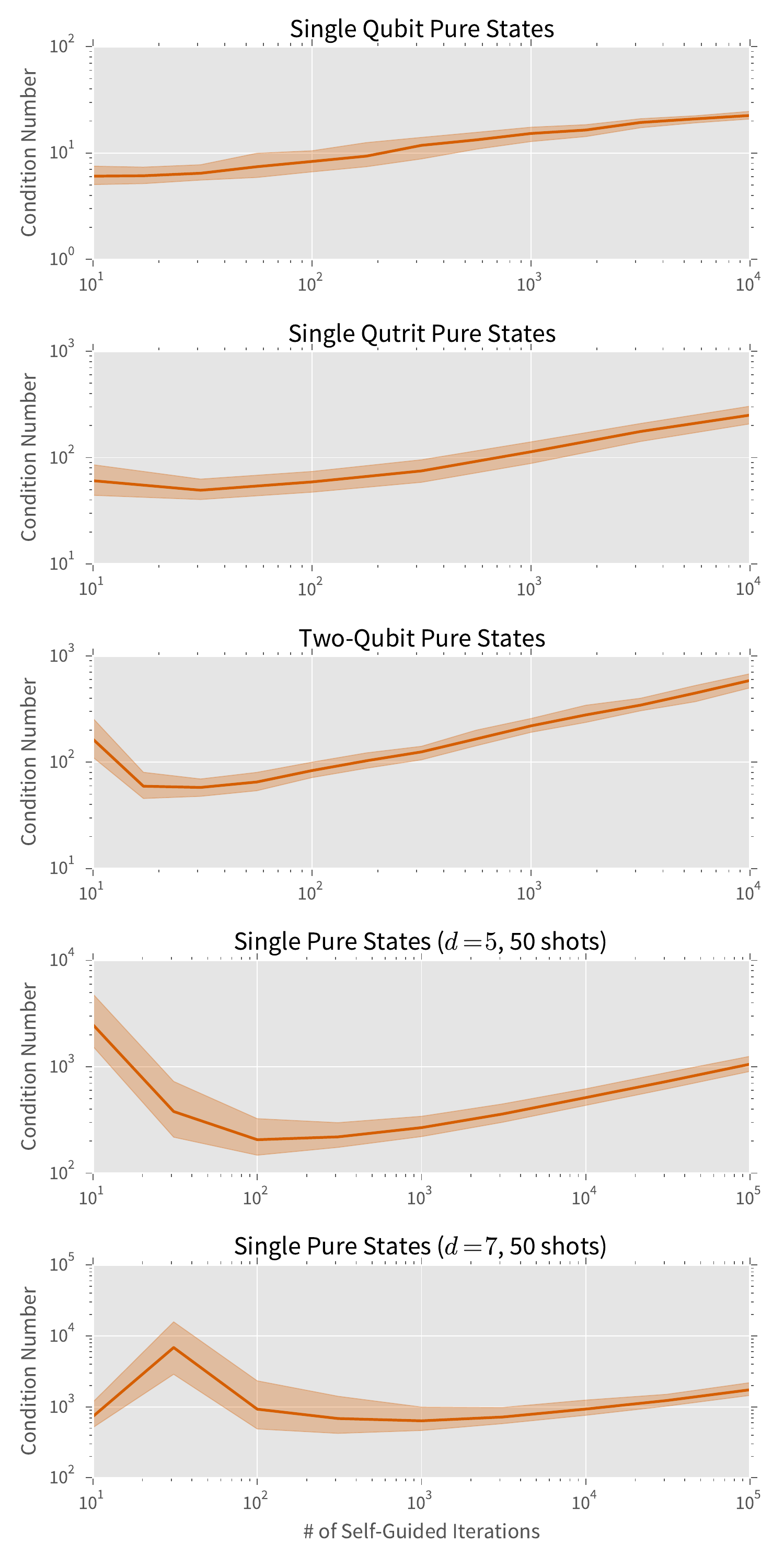}
    \end{centering}
    \caption{
        \label{fig:cond-numbers}
        Condition numbers for least-squares fitting matrices in the single
        qubit, single qutrit, and two-qubit product measurement cases,
        as a function of the number of iterations of self-guided tomography data collection.
    }
\end{figure}

In \autoref{fig:cond-numbers}, we plot the empirical condition number of $\boldsymbol{X}$ as a function of the total number of SGQT iterations. We see the expected behavior. The condition number starts high as there are simply not enough measurements to ensure informational completeness. Then, the condition number reaches a minimum value before rising at a rate of approximately $\sqrt{K}$ due to many nearly (but not exactly) identical measurements \footnote{For the case of $d=7$, the condition number has some interesting transient behavior that we do not yet understand. However, it is still consistent with the asymptotic behavior described above.}.

This effect identifies a fundamental tension between the benefit of measurement adaptivity and offline data analysis, which is why PAQT does well in spite of this tension. 
We note that in most cases using PAQT with a Bayesian mean estimator performs quite well and comes with many added benefits, as discussed above. 
In the cases where the Bayesian mean estimator does not perform well, we conjecture this is due to non-optimal choices of the particle filtering algorithm parameters rather than a fundamental problem of ill-conditionedness. 
This is not a problem to be swept under the rug, however, and a non-trivial optimization will need to be performed to find good operating points for the particle filtering algorithm.

A second comment concerns the standard claim in quantum state tomography work that all results obtained for states will immediately apply to quantum \emph{process} tomography due to the isomorphism between quantum states and channels.
Though this claim is broadly true, there is an important subtlety that we must consider.
Under the Choi-Jami\l{}okowski isomorphism  \cite{choi_completely_1975,jamiolkowski_linear_1972}, process tomography is equivalent to state tomography \emph{with a restriction on allowable priors and measurements}.
Thus, the product-state model of \autoref{fig:two-qubit-infid} is especially important
in that it immediately shows that our adaptive state tomography protocol also
provides a protocol for \emph{process tomography}. Indeed, the Choi-Jami\l{}okowski
isomorphism
gives that product measurements on two copies of a quantum system are equivalent
to preparing a state, evolving under an unknown map, and then measuring the output
state \cite{wood_tensor_2015,granade_practical_2016}.  With this in mind, then,
our results show that self-guided state tomography is an efficient heuristic for
designing quantum process tomography experiments. This will be explored in future work.

\section{Conclusion}

In summary, we have shown how to mitigate the drawbacks of self-guided quantum 
tomography using PAQT.
In numerically testing PAQT, we have shown that SGQT alone is extremely efficient when the true state is pure and 
it is computationally intensive to compete with in higher dimensions. However, more work 
needs to be done to refine the heuristic for mixed states and restricted measurement 
scenarios. We expect that designing good heuristics for the challenging estimation problems 
which lie ahead for quantum technology will become an active area of research, as it has for 
classical machine learning problems.


\begin{acknowledgments}
This work was supported by the US Army Research Office grant numbers W911NF-14-1-0098 and W911NF-14-1-0103, and by the Australian Research Council Centre of Excellence for Engineered Quantum Systems. 
STF acknowledges support from an Australian Research Council Future Fellowship FT130101744. 
We thank Sarah Kaiser for helpful comments.
We thank \citet{okabe_color_2002} for their suggestion of a colorblind-safe
palette for figures and plots.
We acknowledge Thai La Ong restaurant for forgetting about CF's tofu laksa, which forced us to spend an extra 30 minutes at the restaurant and led to a discussion between the authors on the feasibility of the result. 
CG thanks Jacob Bridgeman for assistance in using cluster resources.
\end{acknowledgments}
\newpage

\bibliography{draft_bib,supp_mat}


\appendix
\onecolumngrid

\end{document}